\documentclass{elsart3p}

\usepackage{graphicx}

\usepackage{amssymb}

\usepackage{lineno}

\newcommand{\rmi}{\mathrm{i}}

\begin{document}

\begin{frontmatter}



\title{New aspects of microwave properties of Nb in the mixed state}


\author[RomaTre]{N. Pompeo\corauthref{cor}},
\corauth[cor]{Corresponding author.}
\ead{\\pompeo@fis.uniroma3.it}
\author[RomaTre,SUPERMAT]{E. Silva},
\author[Sapienza]{S. Sarti},
\author[Salerno]{C. Attanasio},
\author[Salerno]{C. Cirillo}
\address[RomaTre]{Dipartimento di Fisica ``E. Amaldi'' and Unit\`a CNISM,
Universit\`a Roma Tre, Via della Vasca Navale 84, 00146 Roma,
Italy}

\address[SUPERMAT]{Laboratorio Regionale SuperMat, CNR-INFM Salerno, I-84081 Baronissi, Italy}
\address[Sapienza]{Dipartimento di Fisica, Universit\`a "La Sapienza", 00185 Roma, Italy}

\address[Salerno]{Laboratorio Regionale SuperMat, CNR-INFM Salerno and Dipartimento di Fisica ``E. R. Caianiello'', Universit\`a degli Studi di Salerno, I-84081 Baronissi, Salerno, Italy}

\begin{abstract}
We present a study of the frequency dependence of the vortex dynamics in a conventional superconductor. We have employed a swept-frequency, Corbino-disk technique to investigate the temperature (3.6K-$T_c$) and high-field (from $H_{c2}/2$ to $H_{c2}$) microwave complex resistivity in Nb thin (20-40 nm) films as a function of the frequency (1-20 GHz). We have found several previously unnoticed features: (i) a field-dependent depinning frequency in the GHz range; (ii) deviations from the accepted frequency dependence \cite{GR}, that can be ascribed to some kind of vortex creep; (iii) the presence of switching phenomena, reminiscent of vortex instabilities. We discuss the possible origin of the features here reported.
\end{abstract}

\end{frontmatter}

\section{Introduction}
\label{intro}
The microwave response of superconductors in the mixed state is a particularly suitable probe to investigate the short-range vortex dynamics. Historically, vortices in conventional superconductors are thought to follow the well-established Gittleman-Rosenblum predictions \cite{GR}, based on single-vortex response to an alternating current without thermal effects. Surprisingly, there are only a few experimental reports on this behaviour, in particular as a function of frequency, temperature and magnetic field. The advancement on knowledge about high frequency vortex dynamics in the recent years \cite{CC}, triggered by the discovery of high-$T_c$ superconductors, and the revamped interest in conventional superconductors as potential components of devices based on superconductor/ferromagnet multilayers \cite{Ioffe}, stimulates a direct, careful experimental determination of the vortex dynamics in conventional superconductors. 

When vortices are set in motion by time-varying currents, they experience heterogeneous forces which include a viscous drag force, phenomenological representation of vortex motion-induced power dissipation, pinning forces, arising from interaction with material defects and hindering vortex movement, and stochastic forces of thermal origin which promote vortex detachment from pinning sites.
The electrodynamic high frequency response arising from the interaction of the fluxon system with microwave currents has been modeled by many authors \cite{GR,CC,brandt,MStheory}. On very general grounds, all models can be represented through a universal expression for the complex vortex resistivity \cite{universal}:
\begin{equation}
\label{eq:rhovm}
    \rho_{vm}=\rho_{vm,1}+\rmi\rho_{vm,2}=\rho_{ff}\frac{\varepsilon+\rmi\left(\nu/\nu_{0}\right)}{1+\rmi\left(\nu/\nu_{0}\right)}
\end{equation}
\noindent where the resistivity $\rho_{ff}$ represents the free flux flow value, reached in the high frequency limit in which vortices experiences only the dissipative viscous drag, $\nu_{0}$ is a characteristic frequency and the dimensionless parameter $0\leq\varepsilon\leq 1$ is a measure of the weight of creep (depinning events) phenomena.
We adopt the Coffey Clem (CC) model \cite{CC}, in which $\varepsilon=[I_0(u)]^{-2}$ and $\nu_0=\nu_p(1-\varepsilon)^{-1}\left(I_1(u)/I_0(u)\right)$. Here, $u=U_0(T,B)/(2 K_B T)$ is the normalized energy barrier height within the assumption of a uniform periodic pinning potential of height $U_0$, $\nu_p$ is the so-called pinning frequency, which marks the crossover between the vortex pinned motion and the purely dissipative motion, arising at $\nu\gg\nu_p$, $I_n$ is the $n$-order modified Bessel functions of the first kind.
The CC model reverts to the simpler Gittleman and Rosemblum (GR) expression $    \rho_{vm}=\rho_{ff}\left(1+\rmi\left(\nu_p/\nu\right)\right)/\left(1+\left(\nu_p/\nu\right)^2\right)$ \cite{GR} for vanishing creep ($\varepsilon\rightarrow0$), so that $\nu_0\rightarrow\nu_p$. In the latter, the pinning frequency can be immediately derived as $\nu_p/\nu=\rho_{vm,2}/\rho_{vm,1}$. 

The exact determination of the three vortex parameters comes directly from comparison to the measured frequency dependence, while only estimates can be obtained from single-frequency measurements resorting to an accurate analysis \cite{universal}. Neglecting the weight of some of the parameters in order to simplify the analysis may significantly affect the estimates of the parameters. 

In the following, we illustrate microwave measurements performed on Nb thin films by means of the Corbino disk technique, which allows wide band, frequency dependent determinations of the complex resistivity.

\section{Experimental results and discussion}
\label{results}
We measure the swept frequency ($\nu$=1-20 GHz) microwave response of Nb samples, at low and constant input power ($<$0.1 mW). Temperature is varied in the range $3.6\;$K$\leq T\leq T_c$ (stability $<0.001\;$K) through a Helium flow cryostat. A static magnetic field $H_{c2}/2\leq H\leq H_{c2}$ is applied perpendicularly to the sample surface.

The studied Nb thin films have thicknesses $d$=20-40 nm and were grown on 0.5 mm thick, $5\times5$ mm$^2$ square Al$_2$O$_3$ substrates in a ultra high vacuum dc diode magnetron sputtering system ($10^{-8}$ mbar base pressure; $10^{-3}$ mbar sputtering Argon pressure). The fabrication was realized at room temperature. The deposition rate of 0.3 nm/s was controlled by a quartz crystal monitor calibrated by low-angle X-Ray reflectivity measurements performed using a Philips X-Pert MRD high resolution diffractometer. 
The sample thickness was determined by fitting the reflectivity profile of the sample with a simulation curve obtained following the Parrat and Nevot--Croce formalism \cite{parrat,nevot}. The fit revealed the formation of a Nb$_2$O$_5$ oxide layer of the order of 2--3 nm on the top of film which causes a reduction of the effective Nb layer thickness. This in turn results in a depression of $T_c$, which is strongly thickness dependent for low thickness values such as ours \cite{Cirillo}. 
Four contact measurements of the dc resistivity on a typical sample 20 nm thick, similar to the one here studied, yielded a constant normal state $\rho_{n}$=22$\;\mu\Omega$cm for temperatures $T_c<T<30\;$K. The critical temperature $T_{c0}=6.37\;$K, evaluated as the midpoint of the resistive transition.

The microwave response is measured through a Corbino disk \cite{sarti} setup: a swept frequency microwave radiation, generated by a Vector Network Analyser (VNA) system, is fed in a coaxial cable short-circuited on the sample. We use a launcher with a custom-made, spring loaded central pin in order to ensure good contact with the Nb film.

We measured the (complex) reflection coefficient $\Gamma(\nu)$ of the electromagnetic wave impinging upon the sample surface. The modulus $|\Gamma|$, bounded within [0,1], is a good indication of power dissipation of the (super)conducting sample. In particular, a vanishingly low dissipation (as in the Meissner state) would yield  $|\Gamma|=1$, whereas by increasing dissipation $|\Gamma|$ would decrease below 1. A quantitative analysis requires the extraction of the complex effective surface impedance $Z_{eff}(\nu)$ of the sample according to the well-known expression \cite{collin}:
\begin{equation}
\label{eq:Gamma}
    Z_{eff}(\nu)=Z_0\frac{1+\Gamma(\nu)}{1-\Gamma(\nu)}
\end{equation}
\noindent where $Z_0$ is the characteristic impedance of the coaxial line. 
The small thickness of our samples allows the use of the thin film approximation in which $Z_{eff}=\tilde\rho/d$, where $\tilde\rho$ is the complex resistivity of the superconductor. We did not observe, in the [$T$, $H$, $\nu$] region here explored, resonances such as the well known substrate resonances occurring in thin films \cite{reson}.

The actually measured reflection coefficient $\Gamma_m(\nu)$, determined by the VNA at its input, include the sample response $\Gamma(\nu)$ as well as the interposed coaxial line response. The calibration of the line contribution is a critical issue, as described in Ref. \cite{sarti}. Here, we used as reference the normal state so that all results are presented as normalized resistivities $\tilde\rho/\rho_n$. 

\begin{figure}[h]
\centerline{\includegraphics{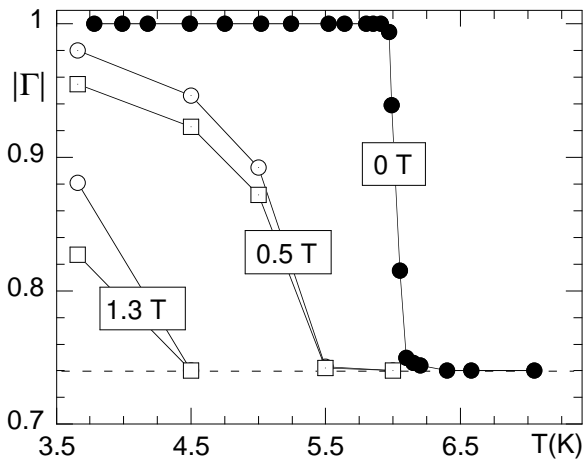}}
  \caption{$|\Gamma|$ vs $T$ at selected fields and frequencies, sample Nb20. Circles: $\nu_1=5\;$GHz; squares: $\nu_2=17\;$GHz.}
\label{fig:trans}
\end{figure}

In the following we consider a Nb sample 20 nm thick (Nb20).  
Raw data are presented in Fig. \ref{fig:trans} to demonstrate the effect of a magnetic field and of the measuring frequency on the superconducting transition.
The black circles correspond to single frequency data taken at $\nu_1=5\;$GHz and $\mu_0 H=0\;$T at various $T$. A finite width of the transition is due to finite quasiparticle contribution that can be observed at high frequency.

The application of a magnetic field (white circles in Fig. \ref{fig:trans}) determines an evident and progressive broadening of the transition, signature of the increased dissipation caused by the motion of the vortices penetrated in the sample. 

In a magnetic field, dissipation increases at larger frequency ($\nu_2=17\;$GHz, squares in Fig. \ref{fig:trans}), directly showing that the vortex characteristic frequency lies in our measuring frequency range.

If, as customarily performed, we rely on single frequency measurements and derive the characteristic frequency neglecting creep (GR model, Ref. \cite{GR}),\footnote{$\nu_p$ can be computed directly from the complex $\Gamma$, given the straightforward derivation $\rho_{vm,2}(\nu))/\rho_{vm,1}=2|\Gamma|\sin(\phi)/(1-|\Gamma|^2)$ where $\phi=\arg(\Gamma)$.}
we obtain different values by using data at different frequencies: 2.1 GHz using data at $\nu_1$, and 3.8 GHz using data at $\nu_2$. Clearly, such a procedure is unsuitable, and wide band measurements are required.

\begin{figure}[h]
\centerline{\includegraphics{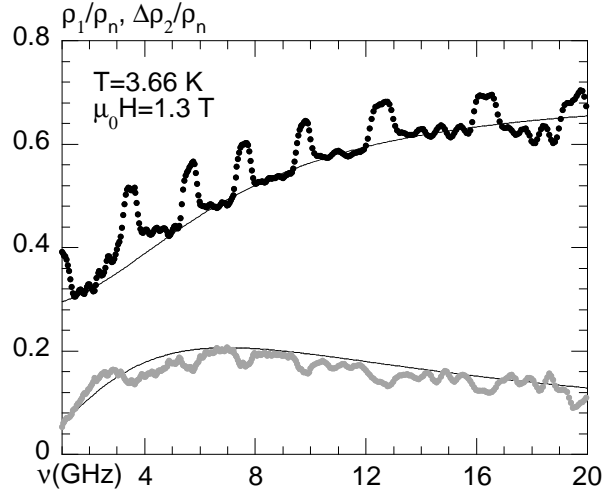}}
  \caption{Normalized $\rho_1$ and $\Delta\rho_2$ (black and grey symbols, respectively) vs $\nu$, sample Nb20. Thin continuous line: fit according to CC model (see text).}
\label{fig:rhovm}
\end{figure}

A sample measurement in the full 1-20 GHz range is reported in Fig. \ref{fig:rhovm}, taken at $T=3.66\;$K and $\mu_0 H=1.3\;$T.
One first, striking, feature is the presence of switching-like phenomena appearing simultaneously on both the real, $\rho_1$, and imaginary, $\Delta\rho_2=\rho_2(H)-\rho_2(0)$, parts of the complex resistivity.
Most likely the phenomenon originates from vortex instabilities, as we will discuss in the following. Here we only stress that it can be thought as superimposing to an underlying, unperturbed response, represented by the lower/upper envelopes of the real and imaginary parts, respectively, of the measured $\tilde\rho$. 
The measured (unperturbed) $\rho_1+\rmi\Delta\rho_2$ shows no signs of quasi-particle/superfluid pairbreaking contributions (as can be expected not too close to $B_{c2}(T)$): $\rho_1$ increases and then saturates at high $\nu$ whereas $\Delta\rho_2$ presents a maximum. These features are clear signatures of single vortex dynamics, so that it can be safely taken $\rho_1+\rmi\Delta\rho_2=\rho_{vm}$.

Therefore, a fit of $\rho_1+\rmi\Delta\rho_2$ can be done by following the CC model. Fit results are reported as thin continuous lines in Fig. \ref{fig:rhovm}: their quality is remarkable. The obtained fitting parameters are $\rho_{ff}/\rho_n=0.7$, $\nu_p=7.2\;$GHz, $\varepsilon=0.41$. 

It can be seen that the pinning frequency $\nu_p$ differs significantly from the simple GR estimate previously calculated. This fact is connected to the presence of significant creep processes ($\varepsilon=0.41$ upon a theoretical maximum value of 1), which evidently excludes the use of simple, GR-like models. This is an important result of this work: we found that significant creep is present even at low (3.66 K) temperatures. It is worth stressing again that these results are made possible by the use of wide band measurements, which demonstrates to be essential for the comprehensive study of vortex dynamics. 

Moreover, a reliable determination of vortex parameters enables also any successive comment on their eventual field dependence. By increasing the field, we observe an increase of creep and a decrease of $\nu_p$, which can be explained by a weakening of vortex pinning by approaching the $H_{c2}(T)$ line.

Finally, the obtained $\rho_{ff}$ yields, within the Bardeen Stephen model ($\rho_{ff}=\rho_n B/B_{c2}$) \cite{BS}, a value of $B_{c2}\sim1.9\;$T compatible with the rough estimate of $B_{c2}\sim2.2\;$T (taking $B\approx\mu_0 H$), determined as the field at which the superconducting signal vanishes within the sensitivity of our system.

\begin{figure}[h]
\centerline{\includegraphics{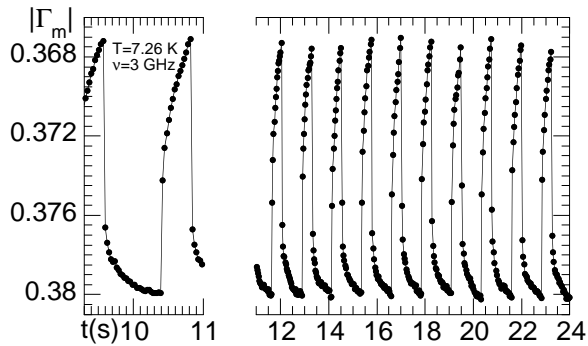}}
  \caption{$|\Gamma_m|$ vs time in residual field $\mu_0 H\sim 30\;$mT, Nb sample 30 nm thick.}
\label{fig:cw}
\end{figure}

We now come back to the switching phenomenon. This is the second main result of this work. Indeed, to the best of our knowledge, this is the first time that similar phenomena are observed in the microwave range.
Our whole set of measurements shows that this phenomenon appears for $H\gtrsim H_{c2}/2$ and increases in amplitude as $H\rightarrow H_{c2}$. Moreover, switches occur upward on $\rho_1$ and downward for $\rho_2$: this feature points to commutations of the vortex system forth to and back from a higher dissipation state.
In order to get additional insight, we performed continuous wave, fixed frequencies measurements. A sample of them is reported in Fig. \ref{fig:cw}, in terms of (uncalibrated) $|\Gamma_m|$ measured as a function of time in a sample $=30\;$nm thick ($T_c\sim7.5\;$K).
The time-dependent nature and the pseudo-periodicity of the switches is clearly apparent, as well as the ``large'' time scales involved (pseudo-period=1.2--1.5 s). Continuous wave data also show no apparent dependence on both the frequency and the power of the microwave stimulus.
Finally, the phenomenon has been observed in all the samples and resulted independent from the instrumentation used.

Although the origin of this switching is at the moment unclear, two main factors are relevant. First, the noted correlation to the vicinity of the $H_{c2}(T)$ line suggests some form of thermally-driven, fluctuation enhanced metastability involving the vortex system \cite{Okuma}. Second, the Corbino disk geometry likely plays a role, since the strong velocity gradients it induces on the fluxon lattice are considered the source of peculiar and unconventional vortex dynamics regimes \cite{Lopez,Lin}. 


\section{Summary}
\label{conc}
We have studied the frequency dependence of the vortex dynamics in Nb thin films at different temperatures and magnetic fields in the range 1-20 GHz through the Corbino disk technique. The resulting single vortex dynamics was well modelized within frameworks such the Coffey Clem one. In particular, all the three main vortex parameters were clearly determined, showing the presence of significant creep contribution even at our lowest (3.66 K) temperatures. In addition we observed time-dependent switching phenomena likely related to metastabilities involving the fluxon system. This phenomenon is probably linked to thermal and fluctuation effects involving the vortex system, as well as to the peculiar fluxon velocity pattern imposed by the Corbino disk geometry.
Further studies will be necessary to ascertain the exact nature of the phenomenon.
\\
\\
This work has been partially supported by an Italian MIUR-PRIN 2007 project.


\begin{thebibliography}{00}




\bibitem{GR} J. Gittleman and B. Rosenblum, Phys. Rev. Lett. {16} (1966) 734.  \bibitem{CC} M.W. Coffey and J.R. Clem, Phys. Rev. Lett. {67} (1991) 386.
\bibitem{Ioffe} L. R. Tagirov, Phys. Rev. Lett. {83} (1999) 2058; B. L. Ioffe et al., Nature { 398} (1999) 697; V. V. Ryazanov et al., Phys. Rev. Lett. {86} (2001) 2427.

\bibitem{brandt} E. H. Brandt, {Phys. Rev. Lett.} {67} (1991) 2219.
\bibitem{MStheory} T. Hocquet et al., Phys. Rev. B {46} (1992) 1061; B. Placais et al., Phys. Rev. B {54} (1996) 13083.
\bibitem{universal} N. Pompeo and E. Silva, Phys. Rev. B {78} (2008) 094503.

\bibitem{parrat} L. G. Parrat, Phys. Rev. { 95} (1954) 359.
\bibitem{nevot} L. Nevot and P. Croce, Rev. Phys. Appl. { 15} (1980) 761.
\bibitem{Cirillo} C. Cirillo et al., Phys. Rev. B 72 (2005) 144511.

\bibitem{sarti} S. Sarti, C. Amabile and E. Silva, arXiv:cond-mat/0406313 (2004).
\bibitem{collin} R. E. Collin {\it Foundation for Microwave
Engineering}, McGraw-Hill International Editions (1992)
\bibitem{reson} E. Silva, M. Lanucara, and R. Marcon, Physica C 276 (1997) 84; N. Pompeo et al., Supercond. Sci. Technol. 20 (2007) 1002.

\bibitem{thin} E. Silva, M. Lanucara, R. Marcon, {Supercond. Sci. Technol.} {9} (1996) 934.
\bibitem{BS} J. Bardeen and M. J. Stephen, Phys. Rev. {140} (1965) A1197.
\bibitem{Okuma} S. Okuma, S. Morishima, and M. Kamada, Phys. Rev. B {76} (2007) 224521.
\bibitem{Lopez} D. L\'{o}pez et al., Phys. Rev. Lett. {82} (1999) 1277.
\bibitem{Lin} N. S. Lin, V. R. Misko, and F. M. Peeters, Phys. Rev. Lett. 102 (2009) 197003.

\end{thebibliography}
\end{document}